# Probing variations of the Rashba spin-orbit coupling at the nanometer scale

Jan Raphael Bindel[1], Mike Pezzotta[1], Jascha Ulrich[2], Marcus Liebmann[1], Eugene Sherman[3], and Markus Morgenstern[1]

[1]II. Institute of Physics B and JARA-FIT, RWTH Aachen University, D-52074 Aachen, Germany

[2] Institute for Quantum Information and JARA-FIT, RWTH Aachen University, D-52074 Aachen, Germany

[3]Department of Physical Chemistry, the University of the Basque Country UPV-EHU and IKERBASQUE, Basque Foundation for Science, Bilbao, Spain

**The Rashba effect as an electrically tunable spin-orbit interaction[1] is the base for a multitude of possible applications[2-4] such as spin filters[3], spin transistors[5,6], and quantum computing using Majorana states in nanowires[7,8]. Moreover, this interaction can determine the spin dephasing[9] and antilocalization phenomena in two dimensions.[10] However, the real space pattern of the Rashba parameter has never been probed, albeit it critically influences, e.g., the more robust spin transistors using the spin helix state[6,11,12] and the otherwise forbidden electron backscattering in topologically protected channels.[13,14]**

**Here, we map this pattern down to nanometer length scales by measuring the spin splitting of the lowest Landau level using scanning tunnelling spectroscopy. We reveal strong fluctuations correlated with the local electrostatic potential for an InSb inversion layer with a large Rashba coefficient (~1 eVÅ).**

**The novel type of Rashba field mapping enables a more comprehensive understanding of the critical fluctuations, which might be decisive towards robust semiconductor-based spintronic devices.**

The Rashba effect[1], which lifts spin degeneracy by breaking inversion symmetry at surfaces or interfaces, was firstly probed in transport using the beating pattern in Shubnikov-de Haas oscillations[15] or the weak antilocalization effect.[16] Later, Rashba–split bands and their spin polarization were visualized by photoelectron spectroscopy.[17] The first successful attempts to use the Rashba effect for spin manipulation required low temperatures and found relatively small signals,[18-21] probably due to D'yakonov-Perel'-type spin randomizing by disorder. Options to overcome this limit are more one-dimensional or ballistic devices.[21,22] Another method is to balance the Rashba and the Dresselhaus couplings,[6] leading to a persistent spin helix with a momentum independent spin rotation axis.[11] For such cases, where the D'yakonov-Perel' mechanism is suppressed, other dephasing mechanisms, such as fluctuations of the Rashba parameter, limit device functionalities. Interestingly, the topological protection in spin channels of 2D topological insulators[13,14] is also likely limited by fluctuations of the Rashba parameter in combination with electron-electron interaction[13] or spin impurities.[14]

A natural method of investigating electronic disorder is scanning tunnelling spectroscopy (STS).[23-24] STS has already revealed fingerprints of the Rashba effect in two-dimensional electron systems (2DES), such as a singularity at the band onset in the density of states (DOS)[25,26], a beating pattern of the Landau levels in the DOS,[27] or standing wave patterns from scattering between Rashba bands and other spin degenerate bands[26], thereby circumventing the absence of quasiparticle interference between the two components of the Rashba-split band.[28] However, these methods did not probe the spatial fluctuations of the Rashba effect.

Here, we use an InSb 2DES, produced by Cs adsorbates on the (110) surface, to probe the Rashba parameter $\alpha$ in real space. STS in magnetic field $B$ reveals a nonlinear spin splitting of the Landau levels (LL), which fits to the Rashba model at intermediate $B$ = 3-7 T, while exchange enhancement[29,30] dominates at higher $B$. Thus, the spin splitting at intermediate $B$ can be used to trace $\alpha$ as a function of position $\boldsymbol{R}$ revealing that $\alpha(\boldsymbol{R})$ spatially fluctuates between 0.4 eVÅ and 1.6 eVÅ. It exhibits a correlation length of 30 nm and a strong correlation with the electrostatic potential of the 2DES, as mapped as the spin-averaged LL energy.

The sample is sketched in Fig. 1a. By adsorbing Cs on p-type InSb(110), the valence and conduction bands are bent down towards the surface such that an inversion layer with one occupied subband is formed (Fig.1b).[27] The Cs coverage (1.8% of a monolayer) is low enough to allow a mapping of the 2DES by STS.[27,29,31] A strong electric field $|E| \approx 3 \cdot 10^7$ V/m within the 2DES results from acceptor-doping,[27] which in combination with the large atomic numbers of In and Sb leads to a large $\alpha$. Figure 1c shows the spin-split LLs of the 2DES according to the Bychkov-Rashba model.[1] One recognizes crossing points of opposite spin levels (dashed ellipses) and a nonlinearity of the spin splitting at low $B$. Figure 1d shows this splitting for different $\alpha$, while keeping all other parameters identical. Different $\alpha$ naturally lead to different nonlinearities, offering an elegant method to locally determine $\alpha$. Although $\alpha$ is a strictly local parameter, the measured spin splitting is related to wave functions, such that the spatial resolution of the method is limited by about the cyclotron radius being smallest for the lowest level $LL_0$. For smooth disorder potentials $V(\boldsymbol{R})$ with respect to the magnetic length $l_B = \sqrt{\hbar/eB}$ (cyclotron radius of $LL_0$), perturbation theory in $\boldsymbol{\nabla} V(\boldsymbol{R}) l_B/(\hbar\omega_c)$ describes the energies $\varepsilon_{n,\lambda}$ adequately for different LLs $n$ and spin labels $\lambda = +,-$:[32]

$$\varepsilon_{n,\lambda}(\boldsymbol{R}) = \hbar\omega_c \left( n - \frac{\lambda}{2} \sqrt{\left(1 - \frac{gm^*}{2m_e} + \frac{V_n(\boldsymbol{R}) - V_{n-1}(\boldsymbol{R})}{\hbar\omega_c}\right)^2 + n\left(\frac{2\sqrt{2}\alpha(\boldsymbol{R})}{\hbar\omega_c l_B}\right)^2} \right) +$$

$$\frac{V_{n-1}(\boldsymbol{R}) + V_n(\boldsymbol{R})}{2}$$

[1]

$$V_n(\boldsymbol{R}) = \int V(\boldsymbol{R}+\boldsymbol{r}) \cdot F_n(\boldsymbol{r}) d^2 r$$

[2]

$F_n(\boldsymbol{r})$ is the kernel of the $n^{th}$ LL wave function (supplement S1) and $\omega_c = eB/m^*$ is the cyclotron frequency. We can therefore determine $\alpha(\boldsymbol{R})$ from the measured splitting $\varepsilon_{0,-} - \varepsilon_{1,+}$, for known $V(\boldsymbol{R})$, $g$ and $m^*$.

Figure 2a shows the DOS, i.e. the spatially averaged local density of states (LDOS) of the 2DES at $B = 7$ T. The characteristic beating pattern of the LLs found previously[27] can be used to estimate the average Rashba parameter $\bar{\alpha}$ by comparison with the different fitting lines. The best agreement with the experimental beating pattern is found for $\bar{\alpha} = 0.7$ eV·Å in agreement with previous results.[27] For simplicity, we used a constant $m^* = 0.03 \cdot m_e$ and $g = -21$ (averaged value, see below), neglecting the band nonparabolicity for the lowest LL. This causes the discrepancies at higher LLs. The width of the LL peaks is directly taken from the distribution of $V(\mathbf{R})$ (Fig. 3e). The observed strong dip at the Fermi level $E_F$ in the experiment is related to the well-known Coulomb gap.[29,33]

In order to extract the local Rashba parameter $\alpha(\mathbf{R})$, we recorded local LL fans. Figure 2b shows the measured LDOS of a single point at different energies and $B$. LLs and spin levels of two subbands are discernible as marked. The individual levels collectively undulate with $B$, which we ascribe to the undulation of all LLs with respect to $E_F$ in order to maintain the fixed carrier density and, to a lesser extent, to exchange enhancement.[29] Reproducible instabilities in the spectroscopy are observed at distinct $B$ (crosses, supplement Fig. S1c). Here, the conductance at $E_F$ partly drops to 3 pS, i.e. an insulating sample area close to integer filling factors prohibits current flow at these $B$. We ascribe the slight suppression of LDOS around $E_F$ to a Coulomb gap.[29,33]

Multiple crossings of levels are present, e.g. in the boxes marked I-III enlarged in Fig. 2c. The dashed lines (guides to the eye) reveal that the marked levels cross away from $B = 0$ T, such that they cannot belong to simple Landau and spin energies, both being linear in $B$ and crossing at $B = 0$ T. A natural way to explain the crossings is the Rashba effect and, indeed, some of the crossings appear at rather similar $B$ as in the calculations of Fig. 1c. Discrepancies, most obvious at lower $B$, are attributed to the local confinement within the potential minimum, where the data are recorded. This complication hampers the use of the crossings for an accurate determination of $\alpha$.

Instead, we use the nonlinearity of the LL$_0$ spin splitting, $\Delta E_{SS} = \varepsilon_{0,-}(B) - \varepsilon_{1,+}(B)$. Figure 2d shows the LDOS recorded at different positions at $B$ = 14 T, exhibiting double peaks for LL$_0$ and more complicated structures for higher LLs. The complex peak structures appear away from the extrema of $V(\mathbf{R})$ due the nodal structure of the LL wave functions.[31] The splitting $\Delta E_{SS}$ determined from fitting two Lorentzians to the two peaks of LL$_0$ is indicated. It increases if the peaks shift to higher energies, i.e., at higher $V(\mathbf{R})$. This is opposite to the expectation from the nonparabolicity of $g(V)$, which decreases with increasing $V$. Furthermore, a fluctuating peak width is observed, which discussion is beyond the scope of this manuscript.

Figure 2e shows $\Delta E_{SS}(B)$ as deduced from Fig. 2b using Lorentzian fits (inset). Above $B$ = 8 T, $\Delta E_{SS}$ oscillates exhibiting maxima at odd filling factors as expected for exchange enhancement.[29,30,33] Since the exchange interaction depends exponentially on the overlap of the wave functions which roughly scales with $l_B \propto \sqrt{1/B}$, it decays rapidly at lower fields, being below 1 meV for $B < 6$ T.[29] Accordingly, oscillations of $\Delta E_{SS}$ are barely discernible at $B < 7$ T. Instead a largely linear $\Delta E_{SS}(B)$ is observed at 3-7 T decaying more rapidly at lower fields, similar to the curves of Fig. 1d. Extrapolating the linear part to $B = 0$ T (dashed line) reveals an offset of $\Delta E_{SS}^0 \sim 2.5$ meV. Taylor expansion for large $B$ of eq. 1 using $V(\mathbf{R}) = const.$, i.e. $V_n(\mathbf{R}) = V(\mathbf{R})$[31] reveals that the offset is given by $\Delta E_{SS}^0(B \to 0) = 4\alpha^2 m^*/(\hbar^2(2 - gm^*/m_e))$ resulting in $\alpha \approx$ 0.65 eVÅ close to $\bar{\alpha} = 0.7$ eVÅ from Fig. 2a. Figure 2f shows the fitted $\Delta E_{SS}(B)$ curve for two different potential minima indicated by crosses in Fig. 3a-d. For the fit, we additionally consider the $V_n$ terms of eq. 1 deduced from the potential $V(\mathbf{R})$ of Fig. 3a (see below). The effective mass is determined from the $B$-field dependent slope of the energy splitting between $LL_0$ and $LL_1$. The $g$-factor then results from the relation $g(V) \cdot m^*(V) = g_0 \cdot m_0^*$, with $g_0 = -51$ and $m_0^* = 0.0135 \cdot m_e$ at the conduction band minimum, leaving $\alpha$ as the only fit parameter.[27] Neglecting the $V_n$ terms would lead to slightly reduced $\alpha$ values ($\alpha(\mathbf{R})$ = 1.04 eVÅ (black cross) & 0.37 eVÅ (red cross)) as has been pointed out previously.[32] More importantly, the obtained $\alpha$ values at those two positions differ by a factor of two indicating strong spatial $\alpha$ fluctuation. Both curves are recorded at potential minima in order to minimize influences by the spatial shape of $V(\mathbf{R})$.

We can rule out the possibility that the spatial fluctuation of $g$ is responsible for the differences, since the curve probed at $V(\mathbf{R})$ = -121 meV (black) should have a larger $g$-factor than that probed at $V(\mathbf{R})$ = -101 meV (red) in contrast to the experimental observation. Exchange enhancement is also much weaker than the obtained differences in $\Delta E_{SS}$, such that a spatially varying $\alpha(\mathbf{R})$ is the most natural explanation.

Having established that $\alpha(\mathbf{R})$ can be deduced from $\Delta E_{SS}$, we produce $\alpha(\mathbf{R})$ maps. A precise determination according to eq. 1 additionally requires $V_n(\mathbf{R})$ maps, estimated by plotting the mean energy of the two $LL_0$ levels (Fig. 3a). The resulting potential fluctuates by about ± 10 meV with a correlation length $l_{VV}$ = 50 nm. Convolving with the LL wave function kernel (eq. 2) leads to the $V_n(\mathbf{R})$ - maps in Fig. 3b, which are obviously smoother than Fig. 3a. Figure 3c shows the $\Delta E_{SS}(\mathbf{R})$ map at 6 T and Fig. 3d shows the resulting $\alpha(\mathbf{R})$ map according to eq. 1. The $\Delta E_{SS}(\mathbf{R})$ and $\alpha(\mathbf{R})$ maps exhibit a similar pattern, but differ in details.

Another contribution to $\alpha(\mathbf{R})$ is the variation of $m^*(R)$ due to the nonparabolcity. We included this in Figure 3d according to $m^*(R, V) = \beta + \gamma \cdot V(R)$ where the effective masses directly deduced from Figs. S2a and b (Supplement) are used as anchor-points to determine $\beta$ and $\gamma$. This barely changes the spatially averaged $\bar{\alpha}_s$ (1 % reduction), but leads to an increase of the FWHM of the distribution of $\alpha(\mathbf{R})$ by 14 %, and a local change of $\alpha(\mathbf{R})$ of even up to 30 %. Notably, $\alpha(\mathbf{R})$ fluctuates between 0.4 eVÅ and 1.6 eVÅ, i.e. by a factor of four. It exhibits a giant $\bar{\alpha}_s = $ 1.2 eVÅ, a rms fluctuation $\delta\alpha = 0.15$ eVÅ (FWHM 30%) (Fig. 3e), and a correlation length of $l_{\alpha\alpha}$ = 30 nm being larger than $l_B$, but smaller than $l_{vv}$ (supplement S5). Therefore, $\alpha(\mathbf{R})$ strongly fluctuates on small length scales.

It is known that $\alpha(\mathbf{R})$ depends mostly on the local electric field $\mathbf{E}(\mathbf{R})$ perpendicular to the 2DES,[1] which is not measurable by STM. However, we find a clear correlation of $\alpha(\mathbf{R})$ with $V(\mathbf{R})$, i.e., a larger $V(\mathbf{R})$ implies a larger $\alpha(\mathbf{R})$ (Fig. 3f). This is expected, since a larger $V$ also implies a stronger local confinement, consequently, a larger $\mathbf{E}(\mathbf{R})$ as sketched in the insets of Fig. 3f. The remaining scatter of $\alpha$ at given $V$ ($\bar{\sigma}_{\bar{\alpha}} = 0.12$ eVÅ), being much larger than the error bar of the $\alpha$ determination ($\bar{\sigma}_\alpha = 0.02$ eVÅ), can be explained by the remaining scatter in the relation between $V$ and $\mathbf{E}$ (supplement S3 & S6).

Finally, we estimate the spin relaxation length $l_{\text{Spin}}$ from the data of Fig. 3d neglecting the D'yakonov-Perel' and the Elliott-Yafet mechanism, i.e., considering only the fluctuations of $\alpha$ for spin dephasing in 2D:[34]

$$l_{\text{Spin}} = \left(\frac{\hbar^2}{2m^*\delta\alpha}\right)^2 \frac{1}{l_{\alpha\alpha}} \approx 270 \, \text{nm}$$

.

The value, not limiting the spin relaxation in this particular system (supplement S8), can take over, e.g. if the D'yakonov-Perel' mechanism is avoided as in ballistic[21,22] or spin-helix[6,11,12] transistors.

Consequently, a detailed understanding of the fluctuations of $\alpha(R)$, as uniquely provided by our novel method, becomes crucial for these prospective devices.

## Methods:

**Preparation of clean InSb(110) surface**

InSb single crystals were glued with a conductive epoxy onto a Mo sample holder. A 1 mm deep notch was cut into the crystal to support cleaving along the (110) surface. A small screw was glued on top of the crystal. Inside an ultrahigh vacuum (UHV) chamber, at a base pressure of $10^{-10}$ mbar, the crystal was cleaved at the notch by pushing the screw towards the chamber wall. After in-situ transfer into a home-built STM within 1 hour and direct cooling to 9 K, 4 K, 1.5 K and 400 mK, respectively, atomically clean and flat terraces with a width of several µm were found.

**2DES inversion layer**

The cleaved surface of the p-doped InSb with acceptor density $N_A = 1 \cdot 10^{24}$ m$^{-3}$ was transferred in UHV into a sample stage hold at $T = 30$ K, and Cs was evaporated from a Cs dispenser onto the surface. The Cs dispenser operated at 470°C contains cesium chromate. After three evaporation cycles of 180 s, the surface coverage is 1.8 % of a monolayer of Cs, defined as one Cs atom per InSb unit cell, as determined by counting the Cs atoms. During the whole procedure, the pressure did not exceed $1.6 \cdot 10^{-9}$ mbar. After the evaporation process, the sample was immediately transferred into the STM and cooled down to 1.5 K.

**Peak fitting for determination of $\varepsilon_{0,-}$, $\varepsilon_{1,+}$, $V(R)$, and spin splitting**

To discriminate the two spin levels of the lowest Landau level, we fit a double Lorentzian peak to the LDOS curves according to

$$\text{LDOS}(E) = \frac{a_1}{\pi} \frac{\sigma_1}{\sigma_1^2 + (E - \varepsilon_{0,-})^2} + \frac{a_2}{\pi} \frac{\sigma_2}{\sigma_2^2 + (E - \varepsilon_{1,+})^2}$$

with amplitudes $a_{1,2}$, peak widths $\sigma_{1,2}$ and peak energies $\varepsilon_{0,-}$ and $\varepsilon_{1,+}$ for peak 1 and 2, respectively. While for B > 3.5 T all six parameters are chosen to be free, for lower $B$, $\sigma_{1,2}$ were both fixed to 5.6 meV in order to deal with the less pronounced spin splitting. This is justified, since the distance of the two levels to $E_\text{F}$ barely changes leading to similar life times and, thus, similar peak widths.

To first order, the LLs probe the electrostatic potential with a resolution of about the cyclotron radius.[32] The electrostatic potential $V(R)$ can, therefore, be estimated as the average of $\varepsilon_{0,-}$ and $\varepsilon_{1,+}$ determined at the position $R$, while the spin splitting is the difference between $\varepsilon_{0,-}$ and $\varepsilon_{1,+}$. Changes in the respective potential and in the spin splitting are not expected on a length scale shorter than $l_\text{B}(6\,\text{T}) = 10.5\,\text{nm}$, hence we smoothed both maps with a Gaussian curve of width $l_\text{B}$.

## Acknowledgements:

We appreciate helpful discussions with D. Hernangómez-Pérez, S. Florens, T. Champel, S. Becker, A. Georgi, C. Saunus, N.M. Freitag, M.M. Glazov, V.K. Dugaev and R. Winkler. Financial support by the German Science foundation via MO 858/11-2 and INST 222/776-1 is gratefully acknowledged. The work of E.S. was supported by the University of the Basque Country UPV/EHU under program UFI 11/55, Spanish Ministry of Economics and Competitiveness (grant FIS2012-36673-C03-01), and "Grupos Consolidados UPV/EHU" program of the Basque Country Government (grant IT472-10). J.U. acknowledges financial support via the Alexander von Humboldt-Stiftung.

## Author contributions:

R. B. prepared the samples, conducted the experiments with the help of M. P. and M. L., evaluated the data with the help of M. L., and wrote the first version of the manuscript together with M. M.. E. S. supported the evaluation of the data and provided the analytic theory to determine Rashba disorder, spin dephasing and spin diffusion length. J. U. provided a theoretical analysis. M. M. and E. S. devised the overall idea of the experiment. All authors contributed to the interpretation of the data and revising the manuscript.

**Figures:**

**Fig.1**

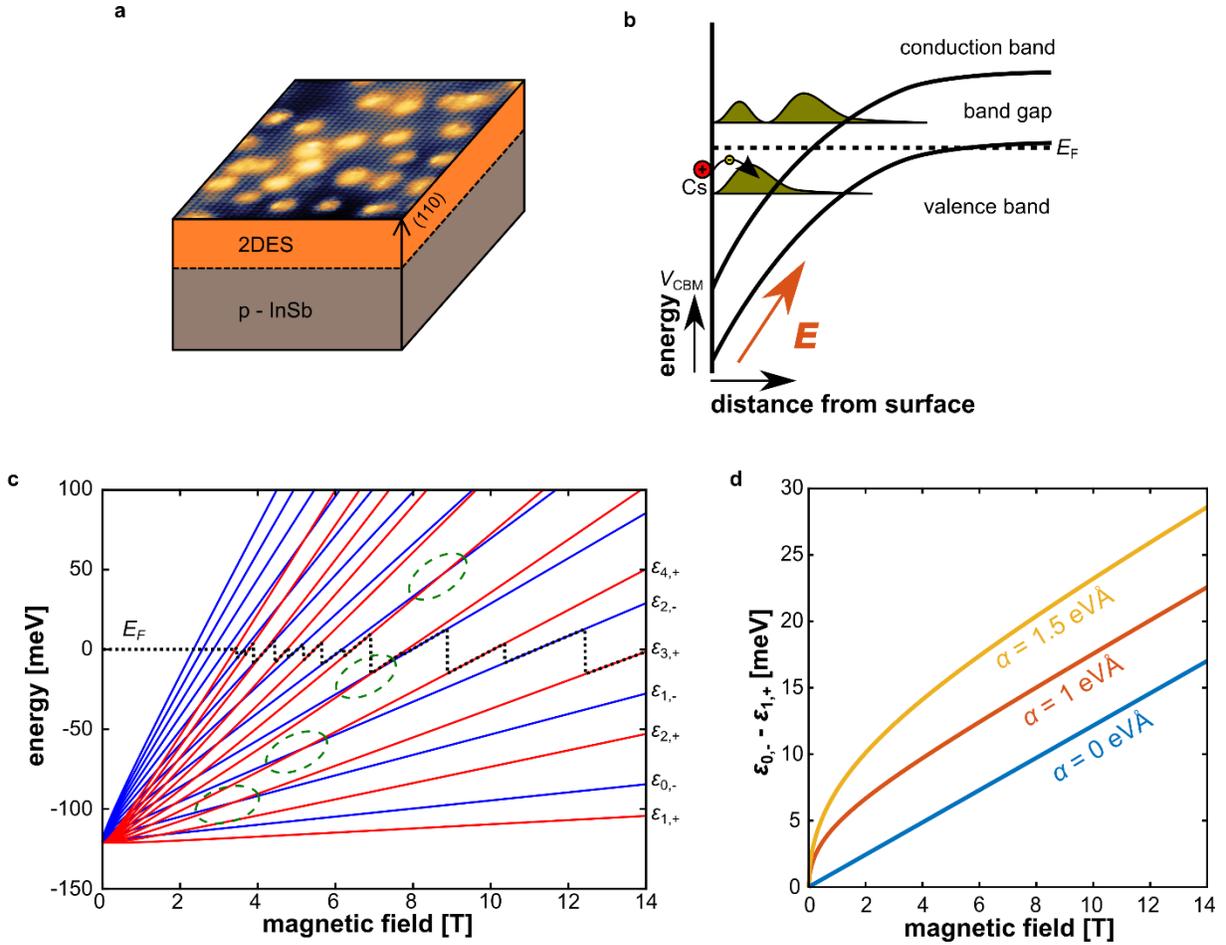

**Figure 1: Rashba parameter from spin splitting of Landau levels. a**, Sketch of the sample with the two-dimensional electron system (2DES) at the surface and conducting p-InSb bulk. An STM image of Cs/InSb(110) with atomic resolution showing Cs atoms (bright dots) on top of the lines of In atoms is displayed on the surface (20 nm × 20 nm, $V$ = 300 mV, $I$ = 30 pA, $T$ = 8 K). **b**, Band structure of the 2DES perpendicular to the surface, as resulting from a Poisson calculation. Confined squared wave functions are sketched in yellow. Adsorbed Cs atoms act as donors. **c**, $B$-field dependence of the energy levels of one 2DES subband using $\alpha = 0.7$ eVÅ, $g = -21$, $m^* = 0.03 \cdot m_e$ (average values from the potential area shown in Fig. 3a). Opposite spin

contributions are marked in blue and red. The Fermi level is shown as a dotted line, using $n = 1.5 \cdot 10^{16}$ m$^{-2}$. Crossing points of different spin levels are highlighted by dashed ellipses. Labels on the right are used in eq. 1. **d**, Splitting of the two lowest energy levels for different Rashba parameters $\alpha$, $g = -21, m^* = 0.03 m_e$.

**Fig.2**

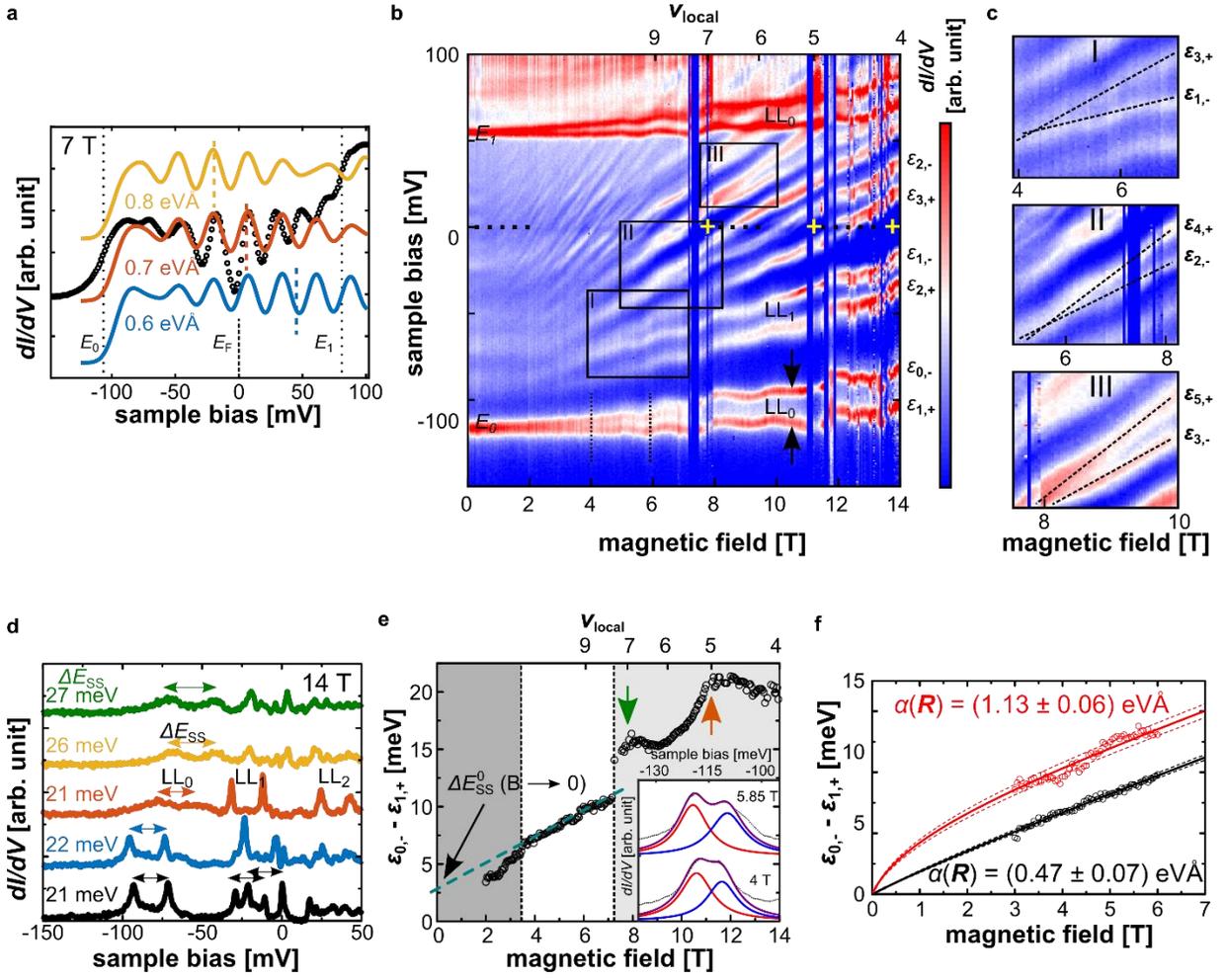

Figure 2: **Local Rashba parameter deduced by STS. a,** Circles: Spatially averaged differential conductance from 150×150 nm² area (35 × 35 pixel) of Cs/InSb(110) ($V_{stab}$ = 300 mV, $I_{stab}$ = 0.2 nA, $V_{mod}$ = 3.5 mV, $T$ = 400 mK, $B$ = 7 T). Dotted lines mark $E_0$ and $E_1$, the onsets of the two different subbands. Full lines: Calculated density of states for different $\alpha$ as marked. Best correspondence of the beating antinode is found for $\alpha \approx 0.7$ eVÅ ($m^*$ = 0.03 $m_e$, $g$ = −21) **b,** $dI/dV$ measurement at a single position within a potential minimum of the 2DES recorded at changing $B$ ($V_{stab}$ = 50 meV, $I_{stab}$ = 0.1 nA, $V_{mod}$ = 0.75 mV, $T$ = 7.5 K). LL$_n$ mark the Landau levels of the lowest subband $E_0$ and the first subband $E_1$, $\varepsilon_{n,\pm}$ mark levels of LL$_n$ according to eq. (1). The local filling factor $\nu_{local}$, marked on top, results from counting all levels up to $E_F$. Signs + mark areas of conductance at $E_F$ below 3 pS. These areas are artificially coloured blue, due to instability (supplement, Fig. S1c). Black arrows mark the spin levels used in (e) and (f). Boxes with roman numbers mark the zoom regions shown in (c). **c,** Zooms into (b) with overlaid guides to the eye (dashed lines), which follow two levels crossing at finite $B$. **d**, Differential conductance recorded at different positions ($V_{stab}$ = 300 meV, $I_{stab}$ = 0.2 nA, $V_{mod}$ = 0.7 mV, $T$ = 400 mK, $B$ = 14 T). Landau levels $LL_n$ and spin splittings $\Delta E_{SS}$, marked by horizontal double arrows, are indicated. The resulting spin splitting of $LL_0$ is additionally marked on the left. Note that multiple peaks appear for higher LLs and that the sharpest $LL_0$ levels are found at the lowest energies. **e**, Circles: energy difference of the two lowest energy levels of (b) (marked by black arrows in (b)). Dashed line extrapolates the nearly linear slope between 3.4 T and 7.2 T towards 0 T as marked by $\Delta E^0_{SS}(B\rightarrow 0)$. Green and orange arrows belong to $\nu_{local}$ = 5 and $\nu_{local}$ = 7, respectively. Inset: $dI/dV$ curves from low energy part of (b) at $B$ as marked (dots) with fit line (violet) consisting of two Lorentzian peaks (red, blue). The energy difference of the maxima of the two Lorentzians is shown as circles in the main image. **f**, Circles: $LL_0$ splitting determined between 3 T and 6 T at two different positions (red, black) as marked by crosses in Fig. 3a-d. Full lines: Fit according to the Rashba model with resulting local Rashba parameter $\alpha(\mathbf{R})$ marked. Dashed lines: 65% confidence interval of the fits with corresponding values of $\pm \alpha(\mathbf{R})$ marked.

# Fig.3

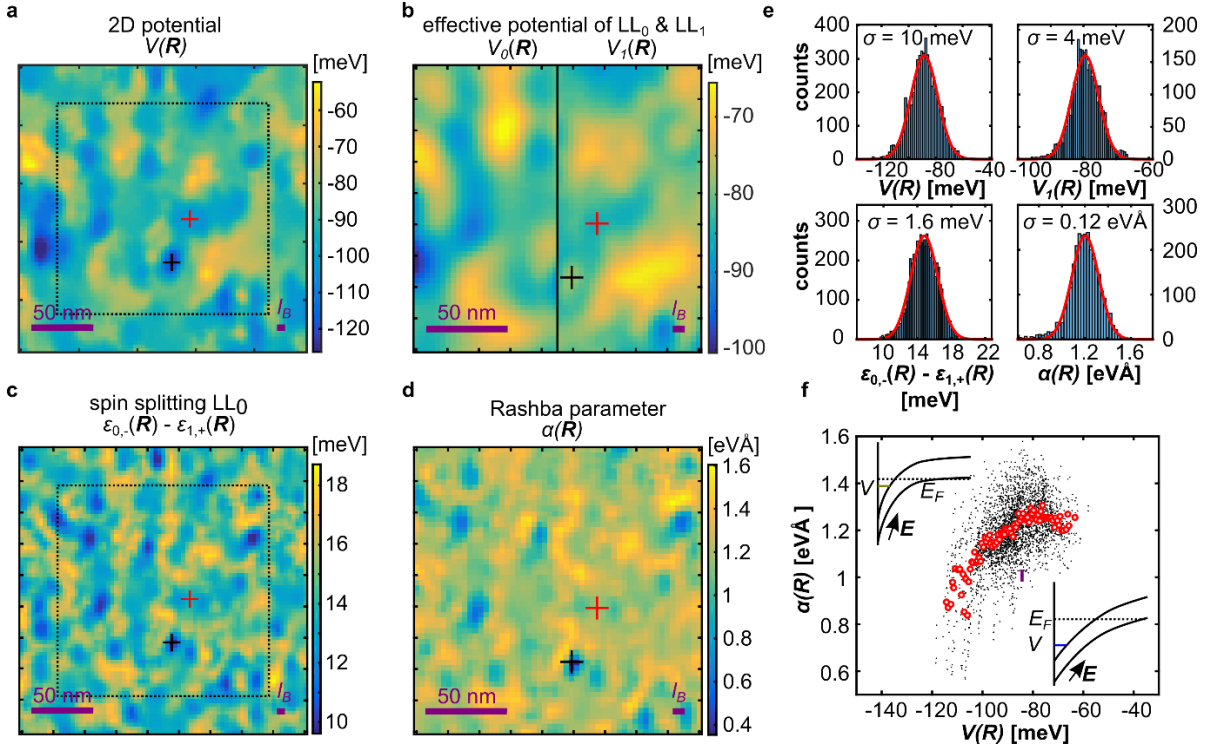

**Figure 3: Mapping of Rashba parameter and comparison with potential map. a,** Map of the potential energy $V(\mathbf{R})$ of the 2DES, which is $(\varepsilon_{0,-}(\mathbf{R}) + \varepsilon_{1,+}(\mathbf{R}))/2$ as resulting from fits of $dI/dV$ data as described in methods and shown in the inset of Fig. 2e ($V_{stab}$ = 50 meV, $I_{stab}$ = 0.1 nA, $V_{mod}$ = 1.5 mV, $T$ = 400 mK, $B$ = 6 T). Black and red cross show positions of curves in Fig. 2f drawn in the same colour. Black dotted square indicates the areas displayed in (b) and (d). The correlation length of the potential is $l_{VV}$ = 50 nm (supplement, Fig. S5c). **b,** $V_0(\mathbf{R})$ (left) and $V_1(\mathbf{R})$ (right) at $B$ = 6 T calculated from the respective area of the $V(\mathbf{R})$ map according to eq. 2. **c,** LL$_0$ splitting ($\Delta E_{SS}$) map at $B$ = 6 T. **d,** Spatially resolved Rashba parameter $\alpha(\mathbf{R})$ determined by using eq. (1), the 2D potential from (a) and the spin splitting map from (c). **e,** histograms of the maps shown in (a), (b), (c) & (d). Gaussian fits (red lines) are added with $\sigma$-values indicated. **f,** Black dots: $\alpha(\mathbf{R})$ from (d) plotted as a function of $V(\mathbf{R})$ from (a) always using the same position $\mathbf{R}$. Red circles: averaged

Rashba parameter across a $V(\boldsymbol{R})$ interval of 0.5 meV. Purple error bar indicates the typical error bars of $V$ and $\alpha$ revealing that the scatter in $\alpha$ is significantly larger than the error bar. Insets: sketch of band bending at high local doping (top left) and low local doping (bottom right) with resulting local 2DES potential $V$ and local $\boldsymbol{E}$-field marked.

# Supplementary information for "Probing variations of the Rashba spin-orbit coupling at the nanometer scale"

Jan Raphael Bindel[1], Mike Pezzotta[1], Jascha Ulrich[2], Marcus Liebmann[1], Eugene Sherman[3], and Markus Morgenstern[1]

[1]II. Institute of Physics B and JARA-FIT, RWTH Aachen University, D-52074 Aachen, Germany

[2] Institute for Quantum Information and JARA-FIT, RWTH Aachen University, D-52074 Aachen, Germany

[3]Department of Physical Chemistry, the University of the Basque Country UPV-EHU and IKERBASQUE, Basque Foundation for Science, Bilbao, Spain

## S1: Kernel functions for the determination of the effective Potential $V_n$

Deducing the Rashba parameter $\alpha(R)$ from the measured spin splitting map requires additional knowledge on the electrostatic potential $V(R)$, which has to be folded by the kernel of the Landau level wave functions (LLWFs) according to eq. 1 and 2 of the main text. To this end, the LLWFs can be rewritten as $\Psi_{n,R}(r)$ with $R = (X, Y)$ being the center of a LLWF and $r = (x, y)$ being the vector from this center to the point of interest[1]:

$$\Psi_{n,R}(r) = \frac{1}{\sqrt{(2\pi l_B^2 n!)}} \left(\frac{z - Z}{\sqrt{2} l_B}\right)^n \exp\left(-\frac{|z|^2 + |Z|^2 - 2Zz^*}{4 l_B^2}\right)$$

Thereby, $Z = X + iY$ and $z = x + iy$ are complex coordinates of the center of the LLWF, often called the guiding center, and of the coordinate relative to this center, respectively. As described in the main text, $l_B$ is the magnetic length and $n$ the LL index.

A straightforward calculation of the related Kernel of the LLWF results in:[1]

$$F_n(r - R) = e^{-(l_B^2/4)\Delta_R} |\Psi_{n,R}(r)|^2$$
$$= \frac{(-1)^n}{\pi l_B^2} L_n\left[\frac{2(r - R)^2}{l_B^2}\right] e^{-(r-R)^2/l_B^2}$$

with $\Delta_R$ being the Laplacian with respect to $R$, $L_n(z)$ the Laguerre polynomial of degree $n$, and $F_{-1}(r - R) \equiv 0$.

A transparent explanation for the inclusion of the resulting $V_n(R)$ into the spin splitting (eq. 1 of the main text) has been provided in reference 1. Due to the Rashba effect, the spin components become mixtures of different Landau levels, which are different for two adjacent spin levels, e.g., for $\varepsilon_{1,+}$ and $\varepsilon_{0,-}$. Since these LLWFs have different lateral extensions, the LLWFs of adjacent spin levels cover different areas of the potential map. Thus, they feel a different average electrostatic potential, which shifts them in energy with respect to each other. Consequently, the spin splitting gets reduced within a potential minimum and enhanced within a potential maximum with respect to the result for a flat lateral potential. For the flat potential, one can show that $V_n(\boldsymbol{R}) = V_m(\boldsymbol{R})$ for all $n$ and $m$, such that the influence of the potential on the spin splitting disappears.

## S2: *dI/dV B*-field sweep

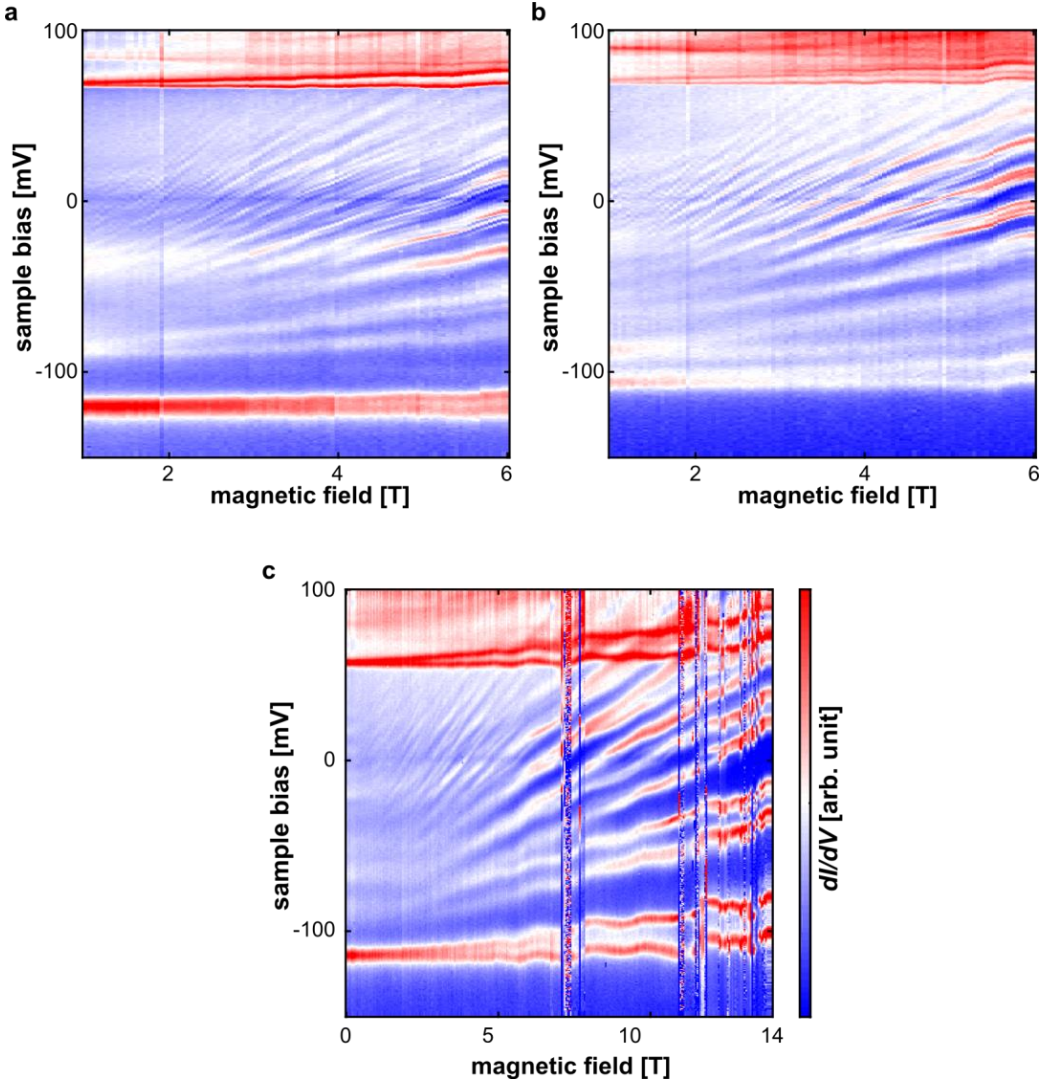

**Figure S1: Mapping Landau level fans at different positions and temperatures.** All *dI/dV(V,B)* spectra are measured within potential minima. The colour code shown at the lower right is valid for all three maps. $V_{stab}$ = 50 meV, $I_{stab}$ = 0.1 nA, $V_{mod}$ = 0.75 mV. **a**, Sweep from 6 T to 1 T, $T$ = 400 mK. **b**, Sweep from 6 T to 1 T, $T$ = 400 mK **c**, Sweep from 0 T to 14 T, $T$ = 7.5 K. This map is the same as the one shown in Fig. 2b of the main text, but without removing the noise appearing at magnetic fields, where the sample resistance gets too large for the STM measurement (noisy stripes).

Figure S1a and S1b show additional *dI/dV(V)* data recorded while slowly ramping the magnetic field from 6 T to 1 T at $T = 400$ mK. In order to ensure that the spectra are recorded at the same position, the recording is interrupted after an increment of one Tesla. Then, an atomically resolved constant-current image is recorded and the tip is readjusted to the same position with respect to the adsorbates visible in the image. This procedure guarantees that a complete LL fan is recorded within 1 nm of the sample surface. The same procedure has been applied in Fig. 2b of the main text.

The resulting LL fan charts again reveal apparent crossings of adjacent levels at finite *B* as well as a suppression of intensity at $E_F$ reminiscent of the Coulomb gap expected for localized systems. While Coulomb gaps have been observed previously for n-type InSb samples, the crossings are exclusive for the current p-type sample with large Rashba coefficient.[2] The spin splitting of $LL_0$ of these maps has been used for Figure 2f of the main text with a (b) related to the black (red) curve. Figure S1c again shows the spectra of Figure 2b, but now without removing the regions of unstable tunnelling conditions.

## S3: Relation between potential *V(R)* and electric field *E(R)*

In Fig. 3f of the main text, it is shown that *α(R)* on average increases monotonously with *V(R)*. However, the scattering of *α(R)* at the same *V(R)* is much larger than the error bar of *α(R)* (see supplement S6). This is related to the fact that *α(R)* is known to be proportional to the electric field |*E(R)*| (Fig. 1b), which is only indirectly represented by the measurable *V(R)* (inset of Fig. 3f, main text). In order to estimate the correlation between electric field and potential, a simple numerical model is applied. Therefore, we distribute positively charged Cs atoms on the surface, which are compensated by randomly distributed negatively charged acceptors within the bulk. Figure S2 shows such a random distribution of negative charges (red dots) with a density of $10^{24}$ m$^{-3}$, which corresponds to the experimental acceptor density. The surface doping is realized by equidistant positive charges (4 nm distance, green dots) mimicking the fact that much more Cs atoms are deposited than electrons are transferred to the InSb, such that the remaining positive charges can arrange rather regularly within the Cs layer.[2]

We simply add up the Coulomb potentials of the positive and negative charges within a plane which is offset by 5 nm from the plane where the acceptors and surface donors are placed. This approximates the average situation of potential fluctuations within a 3D material by removing the measurement plane by about half the inter-acceptor distance from the acceptor plane. As dielectric constants, we use $\epsilon_r = \epsilon_{r,\text{InSb}} = 16.8$ for the negative charges,[3] and $\epsilon_r = 0.5 \cdot \epsilon_{r,\text{InSb}}$ for the positive charges. The latter choice reflects that the Cs atoms are surrounded by vacuum in the upper half-space. Screening itinerant carriers is ignored for the sake of simplicity. The resulting electrostatic potential, respectively, the corresponding electric field in *z*-direction of the inner part of the simulation grid is shown in Fig. S2a. Both, electric field and potential, are finally convolved with the electron probability density distribution of the lowest subband of a triangular potential[4] adapted to the result of the Poisson calculation, with its maximum located at $z_0 = 5$ nm. We checked that the resulting relative fluctuation of the electric field does barely depend on the spatial details of the dielectric constant. For each simulation run, we use only five equidistant *x,y* positions within the inner 50 nm of the measurement plane for the statistical evaluation.

Figure S2b shows the scatter plot of the resulting effective electric fields $|E_{z,\text{eff}}|$ and effective potentials $V_{z,\text{eff}}$ for all evaluated lines. Besides the expected linear dependency of the average electric field on the effective potential (red circles), a significant scattering of the data points is present. The standard deviation of the $|E_{z,\text{eff}}|$ distribution at given $V_{z,\text{eff}}$ is ~10%, which matches nicely to the standard deviation of the Rashba parameter distribution at given $V_{z,\text{eff}}$ shown in Figure 3f of the main text. Therefore, we tentatively assign the observed large scatter of $\alpha(V_{z,\text{eff}})$ to the unavoidable scatter of $|E_{z,\text{eff}}|(V_{z,\text{eff}})$ within a random potential. The absolute value of the effective electric field in our strongly simplified 2D simulation ($\overline{|E_{z,\text{eff}}|} = 1.3 \cdot 10^7$ V/m) is smaller than the electric field of $|E| = 3.1 \cdot 10^7$ V/m resulting from the Poisson-Schrödinger equation, but that does not affect the general conclusion of the unavoidable scatter of $|E_{z,\text{eff}}|(V_{z,\text{eff}})$.[5]

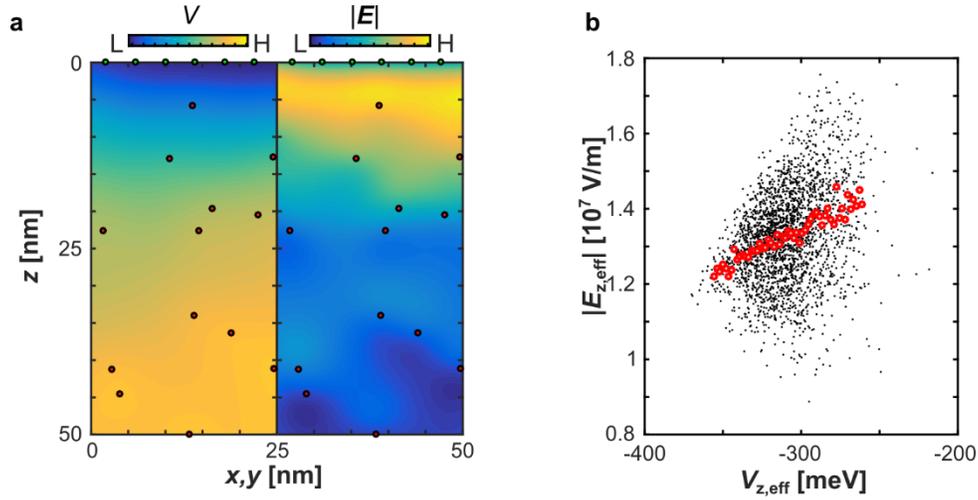

**Figure S2: Interplay between 2DES potential and electric field. a**, Electrostatic potential *V* (left) and electric field in *z*-direction $|E|$ (right) resulting from randomly distributed, negatively charged acceptors (red dots) with density $(\rho_{\text{Zn}} = 10^{24}$ m$^{-3})^{2/3}$, negatively charged surface donors with density $\rho_{\text{Cs}} = 0.25$ nm$^{-1}$ (green dots). The area of simulation is larger than the displayed one (*x* = 150 nm, *z* = 100 nm). **b**, Correlation between effective potential and effective electric field (after folding with the wave function of the first subband of the 2DES) for 2500 simulated different *x*-positions (black dots) resulting from 500 simulations as displayed in (a). Red dots mark the average electric field $E_{z,\text{eff}}$ for a given $V_{z,\text{eff}}$ within an interval of ± 0.25 meV.

## S4: Correlation of the Rashba parameter with lateral gradient and curvature of the electrostatic potential

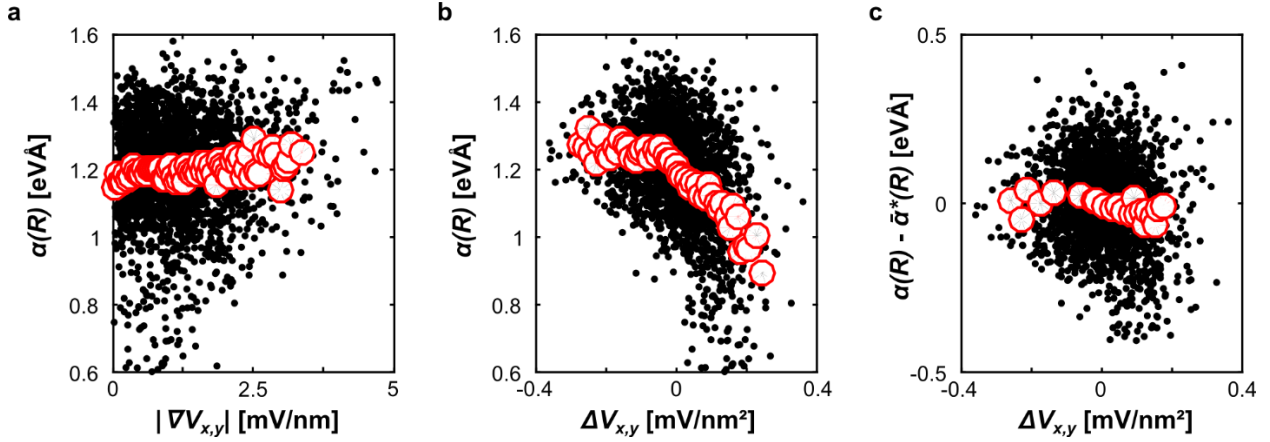

**Figure S3: Correlation of lateral gradient and curvature of the electrostatic potential with the Rashba parameter.** Correlation plots of the experimental data shown in Fig. 3 of the main text. Black dots indicate the data points and red circles the average values. **a**, Correlation between the lateral electric field and the local Rashba parameter. **b**, Correlation between curvature of the 2DES potential and the local Rashba parameter. **c,** Correlation between curvature of the potential and the local Rashba parameter after subtracting the average Rashba parameter found for the corresponding local potential.

In Fig. 3f of the main text, we have shown that $\alpha(\mathbf{R})$ increases monotonously with the electrostatic potential, which is explained by the relation between potential and vertical electric field (Fig. S2b). Due to the presence of the potential disorder, there exists also a lateral electric field $\mathbf{E}_{xy}$, which is the lateral gradient of the local potential $V(\mathbf{R})$. Figure S3a shows the absence of correlation between $\alpha(\mathbf{R})$ and $|\mathbf{E}_{xy}|$. The lateral electric field has negligible influence on the Rashba parameter, which is reasonable, since $|\mathbf{E}_{xy}|$ is about an order of magnitude smaller than the vertical electric fields. In contrast, the curvature of the potential appears to anticorrelate with the Rashba parameter (Fig. S3b). But the curvature itself is not uncorrelated from the potential, i.e. potential minima (maxima) show large positive (negative) curvature. In order to disentangle this indirect effect between curvature and $\alpha(\mathbf{R})$ from a direct influence, we subtract the average value of $\alpha$ found for the particular $V_{z,\text{eff}}(\mathbf{R})$ (red dots in Fig. 3f of the main text) from the measured $\alpha(\mathbf{R})$. The result is shown in Fig. S3c. The correlation obviously disappears. Therefore, as expected, neither the small lateral electric field nor the lateral potential curvature influence the Rashba effect significantly leaving the spatially fluctuating $\mathbf{E}$-field perpendicular to the surface as the central influence on the variable Rashba parameter $\alpha(\mathbf{R})$.

## S5: Correlation lengths of $V(R)$ and $\alpha(R)$

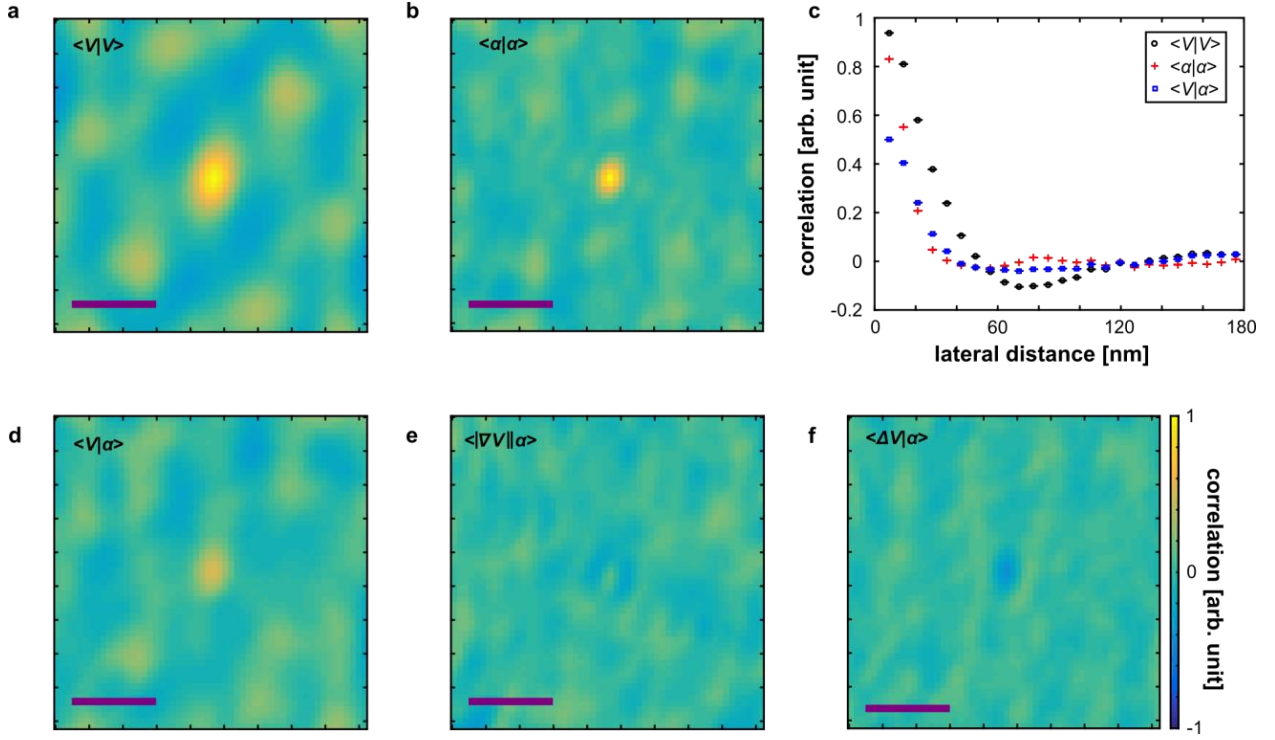

**Figure S4: Auto- and cross-correlations of electrostatic potential map and map of the Rashba parameter.** Correlation maps of the experimental data shown in Fig. 3 of the main text. **a**, **b**, Autocorrelation of the potential map (a) and the Rashba parameter map (b). The correlation length in $\langle\alpha|\alpha\rangle$ is smaller than in $\langle V|V\rangle$. **c**, Radial average of panels (a) (black circles), (b) (red crosses), and (d) (blue squares). The resulting correlation lengths are $L_{VV} = 49$ nm, $L_{\alpha\alpha} = 31$ nm, and are $L_{V\alpha} = 39$ nm. Note that the cross correlation $\langle V|\alpha\rangle$ at 0 nm amounts to ~ 0.6. **d**, **e**, **f**, Cross-correlations of potential and Rashba parameter map, lateral electric field $|\nabla V|$ and Rashba parameter map, as well as potential curvature $\Delta V$ and Rashba parameter map as marked. Scale bars of all panels are 50 nm. Colour code is identical for (a), (b), (d), (e), and (f).

Figure S4 presents different autocorrelation and cross-correlation maps between $\alpha(R)$, $V_{z,\text{eff}}(R)$, and the gradient and curvature of $V_{z,\text{eff}}(R)$. By angularly averaging the correlation maps (Fig. S4c), the correlation length $L_{AB}$ can be determined according to:

$$L_{AB}^2 = 2\pi \int_0^S f_{AB}(x) x \, dx.$$

The upper integration limit $S$ is set as the distance at which the radial averaged function crosses zero for the first time ($S_{VV} = 50$ nm, $S_{\alpha\alpha} = 40$ nm, $S_{V\alpha} = 40$ nm). The dimensionless function $f_{AB}(x)$, with $f_{AB}(0) = 1$, is given by a Gaussian fit to the radial averaged curve up to this point, with the width $\sigma_i$ as the only free parameter. This results in $\sigma_{VV} = (20.1 \pm 0.7)$ nm, $\sigma_{\alpha\alpha} = (12.2 \pm 0.2)$ nm, and $\sigma_{V\alpha} = (16.0 \pm 0.5)$ nm. For the area shown in Fig. 3 of the main text we obtain $L_{VV} = 49$ nm, $L_{\alpha\alpha} = $ 31 nm, and $L_{V\alpha} = 39$ nm, all being larger than the magnetic length $l_B = 10.5$ nm. The fact that $\alpha(R)$ is fluctuating on a shorter length scale than the potential disorder is not completely understood, but might be related to different fluctuation lengths of $V_{z,\text{eff}}(R)$ and $|E_{z,\text{eff}}(R)|$. The cross-correlation map $\langle V(R)|\alpha(R)\rangle$ shows a relatively strong interrelation up to 60%, which becomes, moreover, manifested by the similar features of the real space maps (Fig. 3 main text) and autocorrelation maps of $V(R)$ and $\alpha(R)$.

Cross correlations of $\alpha(R)$, with the gradient and the curvature of $V_{z,\text{eff}}(R)$ are shown in Fig S4e and f. They exhibit rather weak features, which, however, are not analyzed in detail.

Generally, the simultaneous mapping of $\alpha(R)$ and $V_{z,\text{eff}}(R)$ of a 2DES by our novel method opens ample possibilities for a detailed analysis, e.g. via correlation maps.

## S6: Errors on the determination of spin splitting and $\alpha(R)$

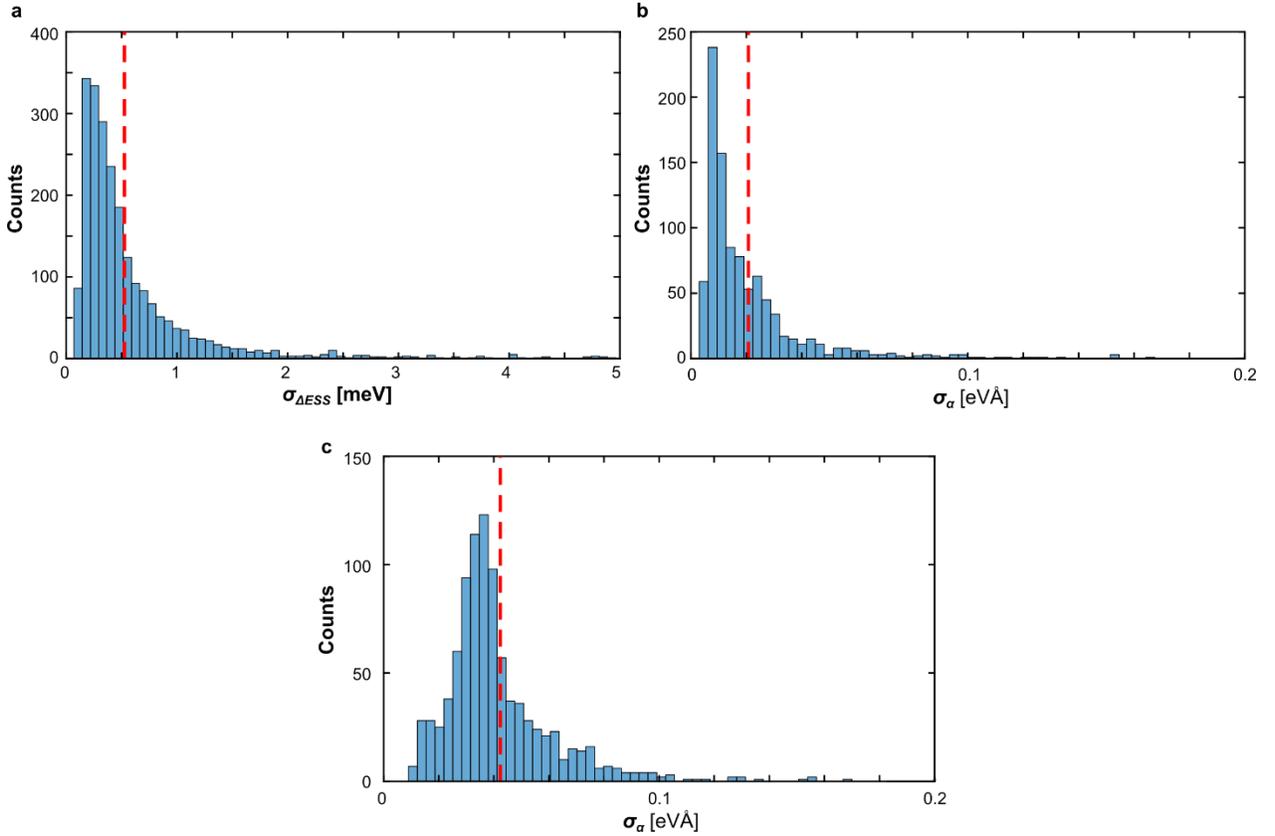

**Figure S5: : Histograms of errors. a**, Error histogram of the spin splitting determination by the fits as shown in Fig. 2e of the main text. The data leading to the spin splitting map in Fig. 3c of the main text are used. 65% of the errors are smaller than $\sigma_{dE} = 0.5$ meV (red line). **b**, error histogram of the resulting local Rashba parameters. 65% of the errors are smaller than $\sigma_\alpha = 0.02$ eVÅ (red line). The error on the determination of $m^*$ (5 %) is not included, since this error results in an uncertainty of $\bar{\alpha}$, but not in uncertainties of the relative $\alpha$–values at different positions, since the same relation between $V$ and $m^*$ is used for the whole image. **c**, error on $α$ with included error on $m^*$. The 65% percentile is $σ_α = 0.04$ eVÅ (red line).

The error on the determination of the spin splitting ($σ_{ΔESS}$) and the local potential value ($σ_V$) is both 0.5 meV (65% percentile). It results directly from the peak fitting of the double Lorentzians (Figure S5a). To determine $α(\mathbf{R})$, firstly, the potential map and the spin splitting map are Gaussian filtered (3x3 pixel, width 10 nm). This reduces the error of $α(\mathbf{R})$ to $σ_α = 0.02$ eVÅ (65% percentile). The corresponding distribution is shown in Fig. S5b.

Due to the nonparabolicity of the dispersion relation, we have to make assumptions for $m^*$, which influence the accuracy of the determination of the absolute value of $α(\mathbf{R})$. We assume that $m^*$ is linear in $E$, which is a simplification, since the evaluated energies relative to the conduction band minimum are larger than the band gap.[6] The linearity is fixed by the two $m^*$, which are determined at two fixed positions with known potential by looking at the $B$-field dependence of the LL splitting $ΔE_{LL}$ at $B$ = 3-6 T ($m^*$ = 0.030·$m_e$ & 0.026·$m_e$). Here, we use $ΔE_{LL} = ℏeB/m^*$ neglecting the contribution of the Rashba effect, which results in an $m^*$ error of 5 %. ($α$ = 1 eVÅ, $m^* ≈ 0.028·m_e$). By this, we estimate the error on the effective mass to $σ_{m^*} = 0.002\ m_e$. This error is an order of magnitude larger than the fitting accuracy of the LL splitting. However, with the inaccuracy of the linear model in mind, we refrain from a recursive more accurate determination of $m^*$. As described in the main text, the local Landé-factor $g$ is determined according to $g(V) \cdot m^*(V) = g_0 \cdot m_0^*$, with $g_0 = -51$ and $m_0^* = 0.0135\ m_e$ increasing the influence of this error additionally.[7,8]

Figure S5c shows the distribution of errors on $α(\mathbf{R})$, taking the uncertainty of $m^*(V)$ and $g(V)$ into account. Thus, the absolute accuracy of $α(\mathbf{R})$ is estimated to $σ_α ≈ 0.04$ eVÅ (4 %), while the relative accuracy at different spatial positions is ~2 %.

## S7: Expected fluctuations of the Rashba coupling due to random acceptors and positive Cs charges

Here we present analytical estimates of the random Rashba coupling and corresponding correlation lengths based on the approach developed in detail in Ref. [9]. In this model, fluctuations in the spin-orbit coupling appear due to random electric fields of the negatively charged acceptors in the depletion layer and the positively charged Cs ions at the surface. As the distribution of the random charge density, we take a similar one to that considered in section S3.

We take the electron probability density distribution (see Fig. 1(b) of the main text) $|Ψ(z)|^2$ as:[4]

$$|Ψ(z)|^2 = (b^3/2) \cdot z^2 \exp(-bz),$$

with the maximum position at $z_0 = 5$ nm, corresponding to $b = 0.4$ nm$^{-1}$. The local value of the spin-orbit coupling is produced by averaging the $z$-component of the electric field over the given probability density $|Ψ(z)|^2$.

As a result of having two sources of the electric field fluctuations, we obtain two correlation functions of the Rashba parameters:

$$⟨δα_{2D}(\mathbf{0})δα_{2D}(\mathbf{R})⟩ = ⟨(δα_{2D})^2⟩F_{2D}(R), \quad ⟨δα_{3D}(\mathbf{0})δα_{3D}(\mathbf{R})⟩ = ⟨(δα_{3D})^2⟩F_{3D}(R)$$

Here, F$_{2D}$ and F$_{3D}$ are the corresponding range functions. They can be calculated straightforwardly using lengthy integral expressions for correlation functions. The subscript 2D is used for the contribution of the surface Cs with the mean density $\bar{n}_{2D}$ and 3D is used for that of the bulk acceptors with the mean density $\bar{n}_{3D}$, and $R$ is the 2D coordinate. We assume here an uncorrelated white-noise distribution of dopants, both, on the surface and in the bulk, described by:

$$\langle n_{2D}(\mathbf{0})n_{2D}(\mathbf{R})\rangle = \bar{n}_{2D}\delta(\mathbf{R}), \quad \langle n_{3D}(\mathbf{0})n_{3D}(\mathbf{r})\rangle = \bar{n}_{3D}\delta(\mathbf{r}), \quad [1]$$

where $r$ is the 3D coordinate. These densities of charged particles produce at a point $r$ a random electric field with unscreened $z$-component [2] in the form

$$E_z(\mathbf{r}) = \frac{1}{\epsilon_{r,\text{InSb}}} e \left[ 2\int n_{2D}(\mathbf{R}) \frac{\mathbf{r}-\mathbf{R}}{|\mathbf{r}-\mathbf{R}|^3} d^2R - \int n_{3D}(\mathbf{r}') \frac{\mathbf{r}-\mathbf{r}'}{|\mathbf{r}-\mathbf{r}'|^3} d^3r' \right]_z,$$

where $d^3r'$ integration is taken over the depletion layer. Analog to supplement S3 we assume that the dielectric constant is $\epsilon_r = 0.5 \cdot \epsilon_{r,\text{InSb}}$ for positive and $\epsilon_r = \epsilon_{r,\text{InSb}}$ for negative charges. Details of the coordinate dependence of the dielectric constant do not have a considerable effect on our results.

By averaging the products $\langle E_z(\mathbf{r}_1)E_z(\mathbf{r}_2)\rangle$ over the disorder by using the distribution $|\Psi(z)|^2$ and the above presented white-noise correlators of the concentrations, one obtains[9], after a lengthy but straightforward calculation, the correlation functions of the random contribution to the Rashba parameters as presented above. The variations of the spin-orbit coupling have the form:

$$\langle (\delta\alpha_{2D})^2 \rangle^{1/2} = (2/\epsilon_r \cdot e^2 \xi) \times b(\pi/10 \cdot \bar{n}_{2D})^{1/2}, \quad [2]$$

$$\langle (\delta\alpha_{3D})^2 \rangle^{1/2} = (1/\epsilon_r \cdot e^2 \xi) \times (\pi/2 \cdot \bar{n}_{3D}b)^{1/2}, \quad [3]$$

where the material parameter $\xi$ describes the proportionality between the electric field and the Rashba parameter.[6] The corresponding correlation lengths are defined similarly to the supplement S5, using the functions $F_{2D}(R)$ and $F_{3D}(R)$ :

$$L^2_{2D(3D)} = 2\pi \int_0^\infty F_{2D(3D)}(R) R dR.$$

This results in:

$$L^2_{2D} = \frac{40\pi}{b^2}, \quad L^2_{3D} = \frac{\pi}{2} \cdot \frac{D_l}{b},$$

where $D_l$ is the depletion layer depth. Note that since the $\delta$-functions of equation [1] do not have a characteristic nonzero spatial scale, the correlation length of the electric fields produced by the surface charges, $L_{2D}$, depends solely on the width of $|\Psi(z)|^2$, that is, $1/b$. However, the bulk dopants are characterized, in addition, by the $D_l$ spatial scale. As a result, $L_{3D}$ includes two length parameters and increases relatively slowly with $D_l$ as the far-distant ions produce relatively weak fluctuations of the electric fields, resulting in the independence of $\langle (\delta\alpha_{3D})^2 \rangle^{1/2}$ of the depletion layer depth in the $D_l b \gg 1$ limit applied to the derivation of equation [2] and corresponding to our system parameters. It is interesting to mention that the correlation lengths depend solely on the system geometry and not on the material and sample parameters such as $\epsilon_r$, $\xi$, and charge densities.

Substituting the characteristic numbers: $\xi = 526$ Å$^2$ (as in Ref. [6]), $\bar{n}_{2D} = 6 \times 10^{16}$ m$^{-2}$, $\bar{n}_{3D} = 10^{24}$ m$^{-3}$, $\epsilon_r = 16.8$ for the dielectric constant of InSb, and $D_l = 30$ nm for the depletion layer width we obtain $\langle(\delta\alpha_{2D})^2\rangle^{1/2} \approx \langle(\delta\alpha_{3D})^2\rangle^{1/2} \approx 0.25$ eVÅ for the variations and $L_{2D} \approx 28$ nm, $L_{3D} \approx 11$ nm for the correlation lengths, respectively. These numbers, being approximate, show a reasonable agreement with the experiment ($\delta\alpha = 0.15$ eVÅ, $L_{\alpha\alpha} = 30$ nm) and allow us to attribute the experimental results to the fluctuations in the electric fields produced by randomness in the distribution of dopants. Notice that the variation in the Rashba parameter due to the acceptors is smeared out in the experiment since $L_{3D}$ is close to the value of the magnetic length.

## S8: Comparison of spin dephasing by fluctuating Rashba parameter and D'yakonov-Perel' mechanism

Here we present the estimates of characteristic values of the spin diffusion length due to the regular and the random spin-orbit couplings of our system. Taking the electron concentration $n = 10^{16}$ m$^{-2}$, we obtain the Fermi wave vector $k_F = \sqrt{2\pi n} = 2.5 \times 10^8$ m$^{-1}$ and the corresponding velocity $v_F = 10^6$ m/s for $m^* = 0.03 \cdot m_e$. For obtained in the main text $\bar{\alpha} = 1$ eVÅ, the regular spin precession rate is $\Omega_{SO} = 2\bar{\alpha}k_F/\hbar \approx 10^{14}$ s$^{-1}$. A typical[10] mobility $\mu = 10^3$ cm$^2$/Vs yields the electron mean free path $l = 15$ nm and momentum relaxation time $\tau = l/v_F = 1.5 \times 10^{-14}$ s. The product $\Omega_{SO}\tau \sim 1$ implies that the spin relaxation, being caused by the electron momentum randomization, is, however, not described by the conventional D'yakonov-Perel' diffusion-like formula, and the spin relaxation time is of the order of the momentum relaxation time $\tau$.[11] As a result, the corresponding spin diffusion length $l_s^{[reg]} = \sqrt{D\tau}$ is close to the free electron path $l$, with $D$ as the diffusion constant given by $D = v_F^2\tau/2$.

The spin-orbit coupling correlation length[7] $l_{\alpha\alpha} = 30$ nm yields for $\delta\alpha = 0.15\bar{\alpha}$ the random contribution to the spin relaxation rate $\Gamma_{rnd} \approx (2\delta\alpha k_F/\hbar)^2 l_{\alpha\alpha}/v_F \sim 2 \times 10^{12}$ s$^{-1}$ and spin relaxation time: $\Gamma_{rnd}^{-1} \sim 5 \times 10^{-13}$ s $\sim 30\tau$. The corresponding spin diffusion length $l_s^{[rnd]} = \sqrt{D\Gamma_{rnd}^{-1}}$ is about $5l_s^{[reg]}$, thus, larger than $l_s^{[reg]}$. Note that the spin relaxation length $l_{Spin}$ as given in the main text describes the traveled path of the electron prior to relaxation and, thus, is larger than $l_s^{[rnd]}$.